\documentclass[aps,pra,twocolumn,amsmath,amssymb,superscriptaddress,showpacs,longbibliography]{revtex4-1}
\usepackage{url} 
\usepackage{amsmath}
\usepackage{amssymb}
\usepackage{amsthm}
\usepackage{fontenc}
\usepackage{graphicx}
\usepackage{xcolor}
\usepackage{textcomp}
\usepackage{epstopdf}
\usepackage{braket}
\usepackage{mathtools}
\usepackage{dcolumn}
\usepackage{multirow}
\usepackage{units}
\usepackage{bbm}
\usepackage{multirow}
\usepackage{array}
\begin{document}


\title{Spin-polarized currents in corrugated graphene nanoribbons}
\author{Hern\' an Santos}
\email[Corresponding author: ]{hernan.santos@urjc.es}
\affiliation{Dept. de Matem\'atica Aplicada, Ciencia e Ingenier\'{\i}a de Materiales y Tecnolog\'{\i}a Electr\'onica, ESCET, Universidad Rey Juan Carlos, C/ Tulip\'an s/n, 28933 M\'ostoles, Madrid, Spain}
\author{A. Latg\' e}
\affiliation{Instituto de F\' isica Universidade Federal Fluminense, Niter\' oi, Av. Litor\^ anea 24210-340, RJ-Brazil}
\author{L. Brey}
\affiliation{Instituto de Ciencia de Materiales de Madrid, Consejo Superior de Investigaciones Cient\'{\i}ficas, C/ Sor Juana In\'es de la Cruz 3, 28049 Madrid, Spain}
\author{Leonor Chico}
\affiliation{Instituto de Ciencia de Materiales de Madrid, Consejo Superior de Investigaciones Cient\'{\i}ficas, C/ Sor Juana In\'es de la Cruz 3, 28049 Madrid, Spain}

\date{\today}

\begin{abstract}
We investigate the production of spin-polarized currents in corrugated graphene nanoribbons.
Such corrugations are modeled as multiple regions with Rashba spin-orbit interactions, where 
concave and convex curvatures are treated as Rashba regions with opposite signs. 
Numerical examples for different  separated Rashba-zone geometries calculated within the tight-binding approximation are provided.
Remarkably, the spin-polarized current in a system with several Rashba areas can be enhanced with respect to the case with a single Rashba part of the same total area. The enhancement is larger for configurations with multiple regions with the same Rashba sign. 
This indicates that the increase of the spin polarization is due to the scattering of the electrons traversing regions with and without  Rashba interaction. 
Additionally, we relate the appearance of the spin-polarized currents
 to novel symmetry relations between the spin-dependent conductances. These symmetries turn out to be a combination of different symmetry operations in real and spin spaces, as those occurring in non-planar systems like carbon nanotubes. 
 Our results show that two-dimensional devices with Rashba spin-orbit interaction can be used as excellent spintronic devices in an all-electrical or mechanical setup.  
\end{abstract}

\maketitle

\section{Introduction}
\begin{center}
\begin{figure*}[t]
\centering
\includegraphics[width=0.85\textwidth]{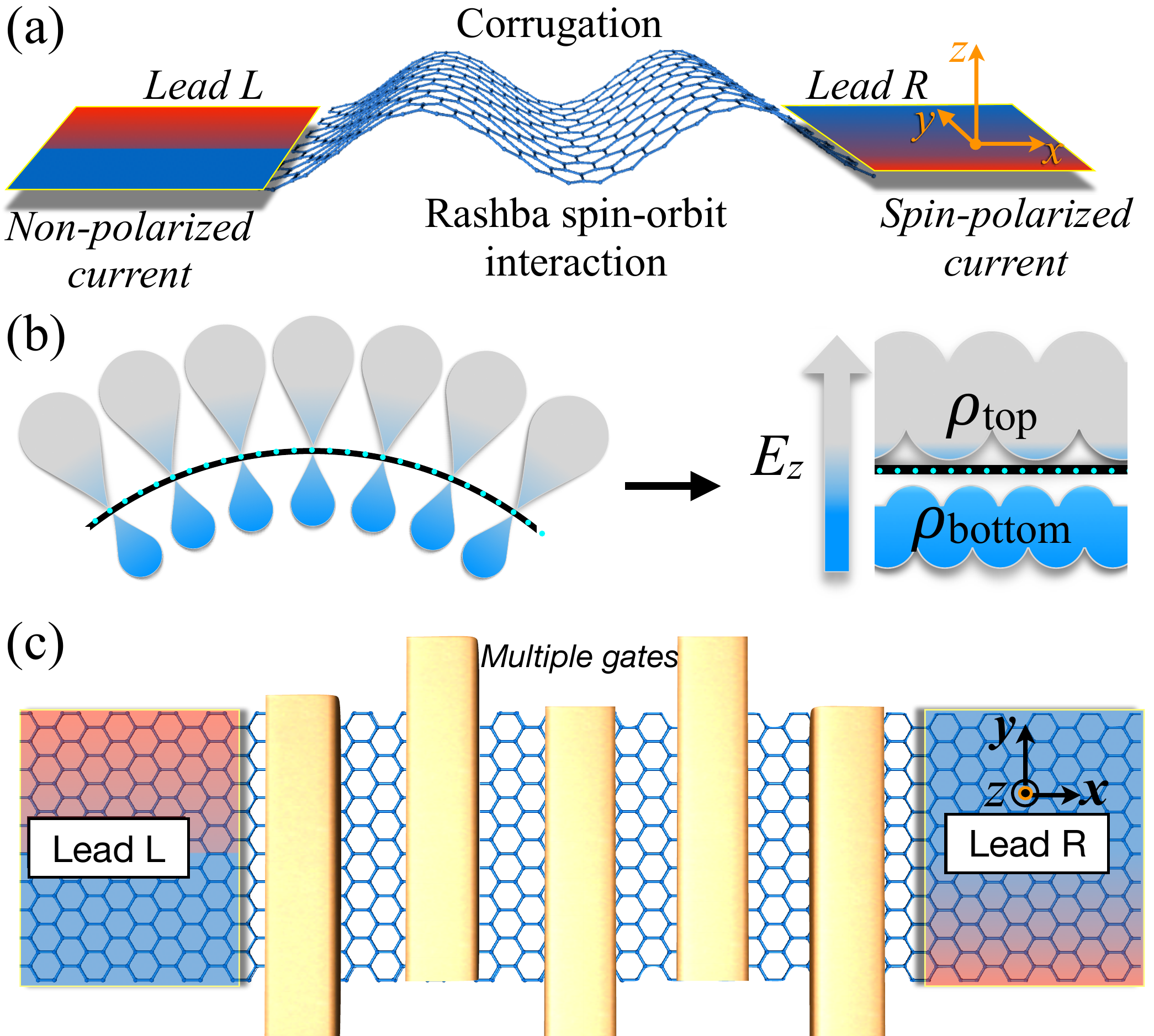}
\caption{\small (Color online)  (a) Schematic device composed by left (L) and right (R) leads and a central region with a corrugated GNR. An unpolarized current  
from the left contact traverses the central part and a spin-polarized current is detected in the right contact. Red and blue are used for the up and down spin components. (b) Schematic $p_z$ orbital distribution in a corrugated graphene in a crest or hill profile. The unbalanced electronic density on the $p_z$ orbitals gives rise to an effective electric field perpendicular to the graphene sheet. (c) The same as (a) but with the central region composed by a multiple-gated graphene nanoribbon. }
\label{fig:1}
\end{figure*}
\end{center}

Corrugated graphene systems have been grown using different experimental techniques and theoretically explored within elastic deformation models, revealing interesting changes in the electronic and transport properties of such systems \cite{Isacsson2008,Deng2016,Bai2014,Androulidakis2017,Costamagna2010,Lim2015}. Lattice deformations produce strain and may modify orbital hybridization; strain effects amount to the appearance of pseudomagnetic fields \cite{Levy2010,Lu2012,Carrillo2014,Torres2018}.  Recently, such corrugations have been achieved by stacking graphene on a self-assembled periodic array of nanospheres;  in such system it was possible to experimentally observe superlattice miniband effects
 \cite{Zhang2018}. Another route to create periodic patterns was recently reported on buckled graphene deposited onto a NbSe$_2$ substrate \cite{Jiang2019} and also in other two-dimensional crystals \cite{Quereda2016,Maeso2019}. The large mismatch between the two materials forces a compressive strain in the graphene membrane, leading to periodic buckled structures whose geometry can be experimentally controlled. The formation of complex mosaic patterns in graphene flakes, by using in-situ uniaxial strain combined with atomic force microscopy, was recently reported \cite{Carbone2019}.  Quantitatively controlled changes in the deformation applied to the sample were achieved, showing that a mechanically tuned device is thus feasible. 
 
Folded graphene sheets have been described as origami-like structures with fascinating properties \cite{Prada2010,Gonzalez2012,Costa2013}.  In fact, 
curved graphene nanoribbons and nanotubes have been shown to present enhanced spin-orbit interactions \cite{Chico2004,Chico2009,Costa2013}. In general, curvature effects 
can produce spin-polarized currents, so corrugated graphene structures, as those depicted in Fig. 1(a), can be used efficiently in spintronic applications. 
From the theoretical viewpoint, a corrugated graphene nanoribbon (GNR) can be described as a superlattice composed of a series of regions with and without an applied electric field, with alternating direction. A structural inversion asymmetry is locally produced by the field, so a Rashba spin-orbit coupling is expected. Curved systems \cite{Chico2012,Santos2013} are known to induce an anisotropic charge distribution in the $p_z$ orbitals due to electronic repulsion, with the subsequent electric field, as schematically shown in Fig. 1(b). In that sense, different corrugation profiles may be explored within this framework, by turning on and off the Rashba spin-orbit interaction (SOI) with the purpose of obtaining spin-dependent currents. Other proposals for the enhancement of spin and valley polarizations have been reported, making use of magnetic barriers on strained graphene with Rashba spin-orbit \cite{Wu2016}. Our scheme does not require the use of magnetic elements, just mechanical deformations or electric fields. 

 On the other hand, periodically repeated regions with Rashba SOI interaction can be induced in planar systems, like graphene nanoribbons, by patterning multiple gates that produce external electric fields perpendicular to the plane of the nanoribbon \cite{Roy2012,Hu2018}. This is schematically shown in Fig. 1(c). 
 Moreover, a recent theoretical work suggests the application of transverse electric fields in twisted ribbons, which is effectively subject to a periodic electrostatic 
potential along its length, with alternating signs \cite{SaizBretin2019}. External gates can be used to tune the transport properties of graphene nanoribbons \cite{Gonza2011}, as well as the value of the SOI coupling.  However, the strength of SOI in graphene systems is very weak, even if curvature or electric fields are used to enhance it \cite{Daniel2006,Min2006}. In particular, the Rashba splitting due to curvature was estimated to be of the order of 0.2 K for curvature radii of 100 nm [28]. Assuming bumps of the order of 1 nm, this could be enhanced to 20 K if such atomic-size corrugations are considered. 
Therefore, other mechanisms should be used to increase the SOI value, such as doping and proximity effects \cite{Cummings2017,Gmitra2017,Wakamura2018,Benitez2018}. 
In fact, a giant spin-Hall effect has been observed in graphene due to the dramatic increase of SOI produced by Cu atoms  \cite{Balakrishnan2014}, reporting spin-orbit splittings around 20 meV. Also, Au intercalation in graphene grown on Ni induces spin-orbit splittings around 100 meV due to hybridization with gold atoms \cite{Marchenko2012}, and in graphene grown on Ir with intercalated Pb atoms, giant values of the spin-orbit splitting were reported
\cite{Calleja2015}. 
Likewise, spin angle-resolved photoemission spectroscopy experiments suggest that a large Rashba-type SOI can be tuned in graphene by the application of an external electric field \cite{Varykhalov2008,Dedkov2008} in samples grown on Ni(111); splittings larger than 100 meV have been measured \cite{Dedkov2008,Zhizhin2015}.

Recently, we showed that depending on their symmetry and the chosen spin projection direction, graphene nanoribbons with a Rashba SOI region
 can yield spin-polarized currents 
 \cite{Chico2015}.  The fact that symmetry reasoning allows to elucidate whether the spin-conserved or spin-flip conductances are equal, can also be used to choose the most suitable ribbons and geometries for spintronic devices. 
We have also explored 
the production of spin-polarized currents in carbon nanotubes (CNTs) by the same physical mechanism, showing that the presence of periodic defects increases its value \cite{Santos2016,Santos2017}. Interestingly, for CNTs the symmetries are more general and it was necessary to consider separately symmetry operations in spin and real space  to account for the relations between spin-resolved conductances. Since the periodic repetition of defects enhances the production of spin-polarized currents, it is natural to explore whether the existence of periodic Rashba regions, arising from structural corrugations or external applied fields via multiple gates patterned on the material, may be a means to augment such effect. We address this issue in the present work.

In this article we study graphene nanoribbons with corrugations, modeled as multiple Rashba regions. We consider either alternating signs of the coupling, which mimics the alternation of concave and convex parts of the ribbon, 
or same-sign Rashba areas, as in bubbles. We start by choosing the spin polarization direction perpendicular to the current and the plane of the ribbon, since this is the optimal geometry to obtain a maximum quantitative effect \cite{Chico2015}. We aim  at elucidating the best configuration to maximize the realization of spin-polarized currents. 

Our main findings are the following: \\
\noindent
(i) Spin-polarized currents in these systems can be enhanced with respect to the case with one single Rashba region of the same size. \\
\noindent
(ii) This effect is larger if the signs of the Rashba regions are the same, which shows that multiple scattering of electrons between regions with and without Rashba SOI is the source of such enhancement. \\
\noindent
(iii) For graphene systems with multiple Rashba regions different spatial and spin symmetry operations have to be considered in certain cases in order to explain the 
relations between spin-resolved conductances and the occurrence of spin polarized currents. These symmetries are different to those obtained previously in carbon nanotubes with Rashba coupling \cite{Santos2016}.  \\
\noindent
(iv) We have performed a symmetry analysis of the Hamiltonians that allows us obtain relations between the spin-resolved conductances depending on the sequence of Rashba regions. 

The paper is organized as follows. Section \ref{sec:model} describes the model employed and the geometries considered. Section \ref{sec:calc} presents numerical calculations that demonstrate the optimal configurations for obtaining the maximum spin-polarized currents. Section \ref{sec:symm} discusses the symmetry issues raised by the numerical results. Finally,  in Section \ref{sec:final} we summarize our conclusions. 

\section{Modeling systems with multiple Rashba regions}
\label{sec:model}
 The proposed device is composed of a corrugated graphene nanoribbon coupled to two leads as shown in Fig. 1(a). The corrugated central part plays the role of a conductor in the transport calculation. Alternatively, the conductor can be a graphene strip with multiple gates, as a truncated superlattice (see Fig. 1(c)). 
 The atomic coordinates of the nanoribbon are those of a perfectly planar system. 
 The corrugations (or the gated area) are described by a sequence of graphene regions perturbed by a Rashba-like SOI due to the electric field, separated by other areas without SOI, that we dub no-Rashba regions. We assume an unpolarized current coming from the left contact traverses the central part; due to the Rashba SOI a spin-polarized current can be detected  in the right contact. 
 
 The whole system is described in the nearest-neigh\-bor hopping tight-binding approximation.  
 The Hamiltonian is given by the sum of a kinetic energy term $H_0=t \sum c_{i}^\dagger c_{j}$, where $t$ is the nearest-neighbor hopping parameter and $ c_{i}^\dagger,  c_{j}$  the destruction and creation operators.
  In the regions with curvature (or a gate potential), the Rashba spin-orbit interaction Hamiltonian is added, \cite{Chico2015,Martino2002,Qiao2010,Lenz2013}, given by
\begin{equation} 
H_R= \frac{i \lambda_R}{a_{ cc}}\sum_{\substack {<i,j>\\\alpha,\beta}} c_{i\alpha}^\dagger  [(\boldsymbol{\sigma}  \times \boldsymbol{d}_{ij}) \cdot \boldsymbol{e_p}]_{\alpha \beta} c_{j\beta} \,\,,
\label{HR} 
\end{equation}
with $\boldsymbol{\sigma}$ being the Pauli spin matrices, $\boldsymbol{d}_{ij}$ the position vector between sites ${i}$ and ${j}$, $a_{cc}$ is the nearest-neighbor carbon-carbon distance in graphene, 1.42 \AA, $\alpha, \beta$ are the spin projection indices, $\lambda_R$ is the Rashba SOI strength due to curvature, which can be additionally tuned by an electric field 
or enhanced by proximity effects, 
and $\boldsymbol{e_p}$  is an unitary vector perpendicular to the plane of the ribbon.
 It defines the electric field direction, either externally applied or appearing due to the corrugation. 
 Note that regions with opposite curvatures are modeled with Rashba terms with opposite signs, maintaining the atoms in a perfect planar geometry. 
 Only nearest-neighbors are considered; previous works in carbon nanotubes have shown that  a four-orbital tight-binding model gives very similar results in the energy range of interest, i.e., near the Fermi level \cite{Santos2016}.

\begin{center}
\begin{figure}
\includegraphics[width=1.0\columnwidth]{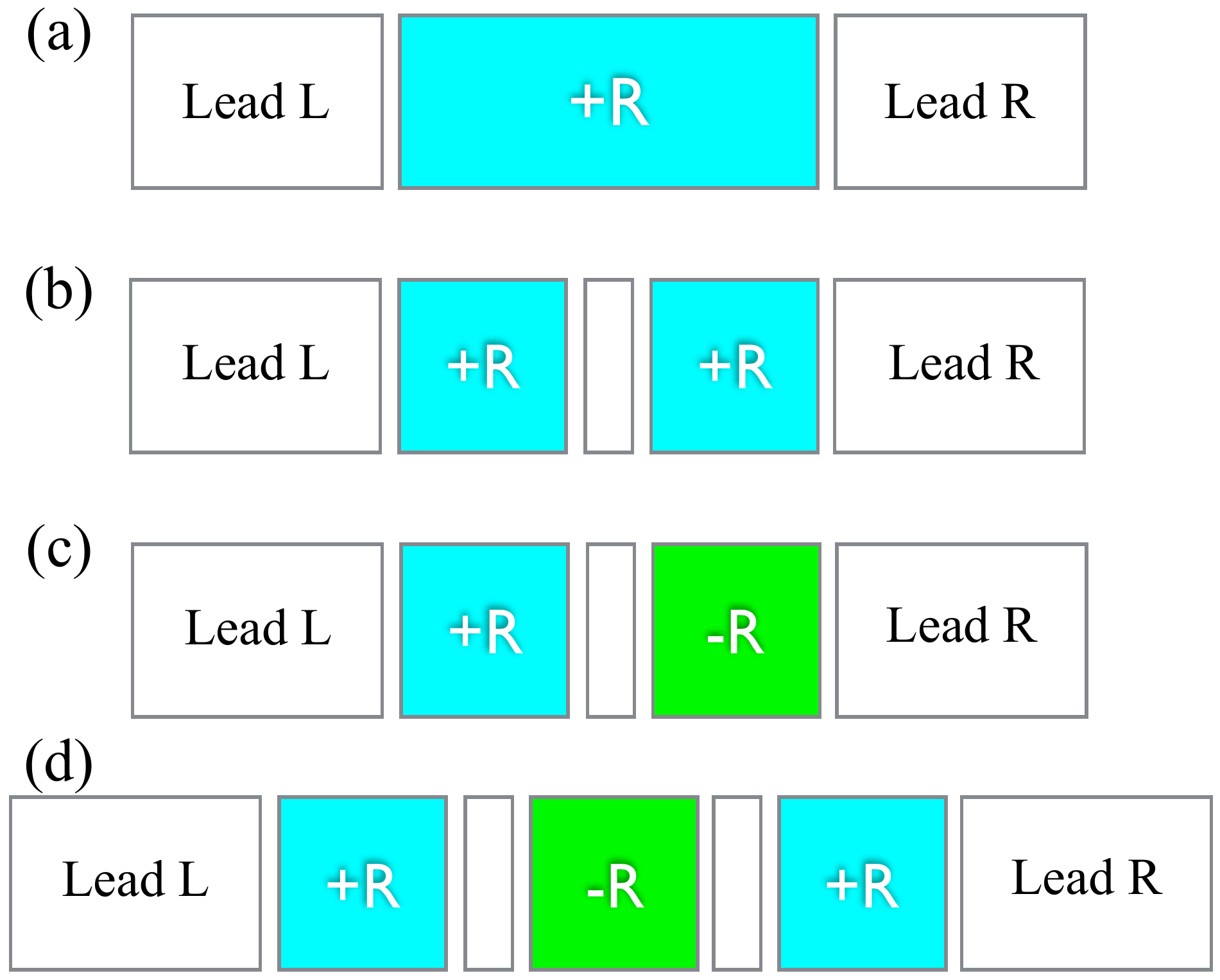}
\caption{\small (Color online)  Schematic view of different corrugated or multiple-gated systems: (a) homogeneous central part; (b) two Rashba regions with the 
same sign of the Rashba coupling separated by a no-Rashba region; (c) the same as (b) with opposite Rashba signs, and (d) three Rashba regions with  alternating sign Rashba coupling, separated by no-Rashba regions.}
\label{fig:2}
\end{figure}
\end{center}

If the corrugated graphene nanoribbon is shaped in a series of hills and valleys, 
it is described by a sequence of positive and negative Rashba coupling regions to take into account the positive and negative concavities produced by the folding. For corrugations with the same curvature, as the bubbles reported in Ref. \cite{Jiang2019}, the folds can be modeled as regions with the same sign of the Rashba SOI coupling. 
 Any of these arrangements can be also achieved with multiple-gated systems, 
 for which the sequence of the Rashba signs is determined by the transversal gates. 
 As in Ref. \cite{Costa2013}, we have not considered  an inhomogeneous variation of the curvature, modeled with a gradual variation of the Rashba coupling in the region with a non-zero electric field. We verified for a previous work in carbon nanotubes with defects \cite{Santos2017}, that an inhomogeneous profile, assuming   a linear dependence of the Rashba coupling at the interface, gave results similar to those adopting an abrupt interface. Moreover, other works on Rashba coupling in quantum wires also show that  spatial variations on the Rashba profiles do not yield significant changes in the spin conductances \cite{Sanchez2008}.
 However, this issue should be also checked for the geometries explored here, namely, graphene nanoribbons with several Rashba regions. 
In the Supplementary Material we show this explicitly for one of the cases studied in this work. 

Some particular examples of the geometries studied are shown in Fig. \ref{fig:2}.  In Fig. \ref{fig:2}(a) the central conductor is chosen as  a single region with positive SOI, denoted here as $+R$; this is the case studied in Ref. \cite{Chico2015}.  Figs. \ref{fig:2}(b) and (c) depict two regions with the same SOI sign (+R,+R) and opposite signs (+R,$-$R), respectively, separated by a region without Rashba coupling. Finally, in Fig. \ref{fig:2}(d) a positive-negative-positive sequence (+R,$-$R,+R) with spacers without SOI in between is depicted.
 
For those systems with the same sign of the Rashba coupling in all regions (Fig. \ref{fig:2}(b)) or with an odd number of alternating Rashba regions (Fig. \ref{fig:2}(d)), the sequence of signs of the Rashba term is the same starting from the left or from the right lead.  
The symmetries of these systems are the same as for the single Rashba region (Fig. \ref{fig:2}(a)) discussed previously \cite{Chico2015,Chico2020}.  
 Symmetries acting simultaneously in real and spin space are sufficient to explain the relations between spin-resolved conductances. 
We call these L-R symmetric systems. However, for systems with an even number of alternating Rashba regions (Fig. \ref{fig:2}(c)),
the sequence of signs of the Rashba regions starting from the right lead is reversed with respect to that obtained starting from the left lead.
  This should be considered in order to find the symmetries of the total Hamiltonian. We will later see that this situation calls for different symmetry operations in real and spin space, as in carbon nanotubes \cite{Santos2016}.
   We denote these systems as L-R antisymmetric.  These require additional symmetries with respect to the nanotube case, which involve the swapping of the  Rashba regions.

The conductance is computed within the Landauer approach by using the Green function formalism \cite{Chico1996,Xu2007,Diniz2012}. 
%
%
The spin-resolved conductance  $\textit{G}^{LR}_{\sigma \sigma'} $ is proportional to the probability that one electron from the left (L) electrode and spin projection $\sigma$ reaches the right (R) electrode with spin projection  $\sigma'$.  
Notice, that for the sake of simplicity, we omit in the conductances the index indicating the spin projection direction $s$, that could be $x, y$ or $z$. Thus, $\sigma$ and $\sigma'$ can take the values $\uparrow$ or $\downarrow$. 
In terms of the Green's functions, $\textit{G}^{LR}_{\sigma \sigma'} $  is given by \cite{Chico2015}
\begin{equation}
\textit{G}^{LR}_{\sigma \sigma'} = \frac{e^2}{h} Tr[\Gamma^{L}_{\sigma}G^r_{\sigma,\sigma^\prime}\Gamma^{R}_{\sigma'}G^a_{\sigma',\sigma}]\,\,,
\end{equation}
where $G^{a(r)}_{\sigma, \sigma^\prime} $ is the advanced (retarded) Green function of the conductor and $\Gamma^{L(R)}_{\sigma}=i[\sum^{r}_{L(R),\sigma} - \sum^{a}_{L(R),\sigma}]$ is written in terms of the $L$ $(R)$ lead self-energies $\Sigma^{a,r}_{L(R),\sigma}$. 
The spin polarization of the current in the $s$ direction is defined as  
\begin{equation} 
P_s=G^{LR}_{\uparrow \uparrow}+G^{LR}_{\downarrow \uparrow}-G^{LR}_{\downarrow \downarrow}-G^{LR}_{\uparrow \downarrow}. 
\label{pol} 
\end{equation}
Notice that this definition \cite{Chico2015,Santos2016,Santos2017}  is most directly related to the magnitude actually measured, which is the current. 
In fact, in the low-bias limit, linear response theory holds and therefore the conductance is proportional to the current. 
Since the spin polarization of the current arises when there are several channels in the leads, we consider this choice more representative than the dimensionless definition adopted by other authors; 
i.e., a normalized value obtained by dividing Eq. \ref{pol} by the total conductance $\sum_{\sigma, \sigma'} G^{LR}_{\sigma \sigma'}$, as done in Refs. \cite{Diniz2012,Zhai2005}. 

Seeing that the spin polarization of the current arises from the difference between the spin-conserved and/or the spin-flip conductances, in what follows we name spin-conserved polarization to the difference between the spin-conserved conductances $G^{LR}_{\uparrow \uparrow}-G^{LR}_{\downarrow \downarrow}$, and the difference $G^{LR}_{\downarrow \uparrow}-G^{LR}_{\uparrow \downarrow}$ is denoted as spin-flip polarization. 
As mentioned above, the spin projection direction that maximizes the effect is transversal, i.e., perpendicular to the current flow and electric field \cite{Chico2015, Santos2016, Santos2017}.  Therefore, all the numerical calculations are done for this spin polarization direction ($y$ in our choice of axes, see Fig. \ref{fig:2}). 
Notwithstanding, the symmetry relations of the conductances will be discussed for all spin projection directions. 
%

The Rashba coupling constant is taken as  $\lambda_R= 0.1t$,  which is a large value for pristine graphene, but 
in the range of first-principles calculations in graphene with absorbed Au atoms \cite{Ma2012}, 
or close to the experimental results for graphene with proximity-enhanced SO  due to decoration with Ni, for which Rashba splittings of 0.225 eV were reported \cite{Dedkov2008}. 
Proximity-enhanced SOI in corrugated graphene could be obtained in experimental setups in which graphene is grown in a Cu or Ni substrate the produces wrinkles, as reported in Refs. \cite{Chae2009,Meng2013}. 
Besides, we expect our conclusions to be valid for other materials such as germanene or silicene \cite{Liu2011,Rzes2018}, especially in proximity to other layered systems with large SO coupling. 
In any case, our conclusions hold for smaller values of the coupling SOI constant, compatible with the SOI splitting of 225 meV reported experimentally in graphene with Ni \cite{Dedkov2008}. We will show this explicitly for one of the systems studied.

\begin{center}
	\begin{figure*}[t]
	\centering
	\includegraphics[width=1.\textwidth]{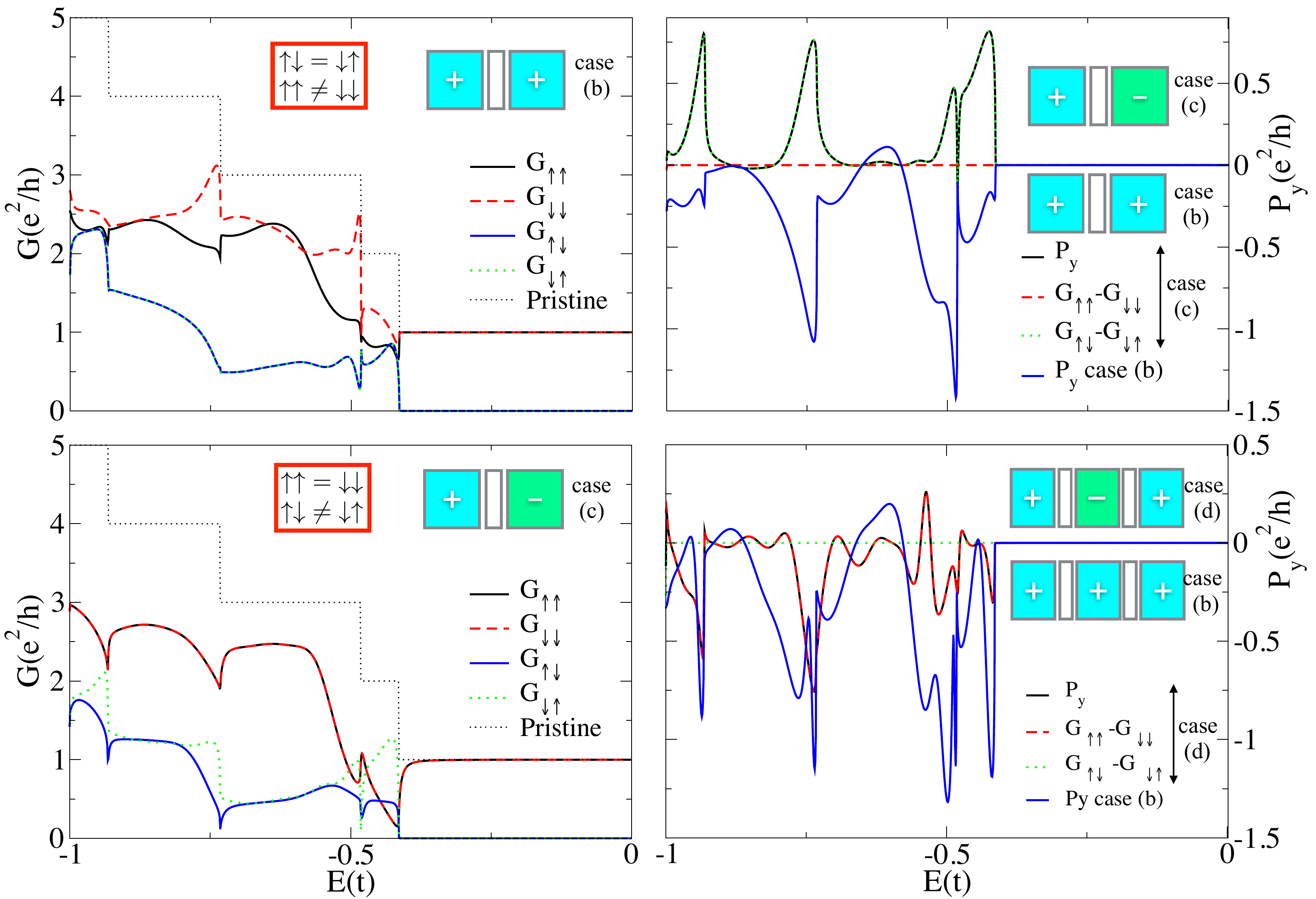}
	\caption{\small (Color online) Left panels: spin-dependent conductances projected on the $y$ direction for 11-AGNR systems composed by 4 unit cells of Rashba region plus an intermediate region without SOI of x=1 unit cell. The top panel shows the (+R,+R) case, whereas the bottom panel is the (+R,$-$R) case.   The spin-resolved conductances $G_{\uparrow \uparrow}$, $G_{\downarrow \downarrow}$, $G_{\uparrow \downarrow}$, and $G_{\downarrow \uparrow}$ are plotted individually. The conductance per spin channel for the pristine 11-AGNR (without Rashba SOI) is depicted in black dotted lines. Right panels:  Spin polarization of the current $P_y$ of several 11-AGNR systems with the same characteristics (4 uc Rashba regions with x=1 no-Rashba spacers). At the top, $P_y$ for the cases depicted in the left panels, namely, (+R,+R) and (+R,$-$R). Bottom: systems with three Rashba regions,  (+R,+R,+R) and (+R,$-$R,+R). }
	\label{fig:3}
	\end{figure*}
\end{center}

\section{Results}
\label{sec:calc}

In what follows, the widths of the armchair and zigzag graphene nanoribbons  (AGNRs and ZGNRs) are given in dimers or chains, respectively \cite{Nakada1996}. 
For the lengths we use the translational unit cell (uc), equal to 3$a_{cc}$ in AGNRs and $\sqrt{3}a_{cc}$ in ZGNRs (see, e.g., Ref. \cite{Chico2015}).

The left panels of Fig. \ref{fig:3} show the spin-dependent conductances for 11-AGNR systems, with the spin projected in the $y$ direction.  Top and bottom panels display the (+R,+R) and (+R,$-$R) configurations, respectively. The systems are composed by 4 uc of Rashba region plus an intermediate region without SOI (x=1 uc). 
 For comparison, the conductance 
per spin channel 
for an 11-AGNR without Rashba coupling is depicted in black dotted lines.
As explicitly depicted in the red square insets, the case (+R,+R) exhibits different spin-conserved conductances ($G_{\uparrow \uparrow}\neq G_{\downarrow \downarrow}$). The situation is completely reversed for the  (+R,$-$R) case, where the conductance differences come from the spin-flip terms. This dissimilar behavior has its origin on the symmetries analyzed in Section 4.

The right panels of Fig. \ref{fig:3}  present the spin polarization of the current, $P_y$, for the cases depicted in the left panel (top) and two related systems, with an extra Rashba region (bottom). 
  Given that electron-hole symmetry holds and therefore $P_s(E) = -P_s(-E)$ \cite{Chico2015}, the energy interval is represented from $-t$ to 0. 
 $P_y$ is nonzero above/below $\pm0.41t$, i.e., at energies for which the second channel in the leads is available for conduction. 
 Since time-reversal symmetry is also preserved, two channels at the outgoing lead are necessary to obtain spin polarization \cite{Zhai2005}. Obviously, it is possible to decrease the energy threshold for the spin polarization by increasing the width of the ribbon, but for the sake of clarity in the figures we stick to narrow ribbons.

The right bottom panel of Fig. \ref{fig:3} shows the spin polarization of the current for the same 11-AGNR but with three Rashba regions: (+R,+R,+R) and (+R,$-$R,+R). Notice that, differently from the case of two Rashba regions with opposite Rashba signs, the $P_y$  corresponding to the (+R,$-$R,+R) configuration arises from the difference between spin-conserved conductances. 
The same happens to the (+R,+R,+R) system; both are L-R symmetric and show the same behavior in this respect. 
We find that  the spin polarization of the current  $P_y$ is larger for systems with the same sign of the Rashba coupling, either with even or odd number of Rashba regions, 
in the sense that the maximum values are obtained for systems with same-sign Rashba regions, although for an specific energy this may not be correct. For larger systems or more scattering interfaces this visual inspection is harder to carry out, so we will introduce below a magnitude to quantify this effect.

\begin{center}
	\begin{figure*}[t]
	\centering
	\includegraphics[width=1.\textwidth]{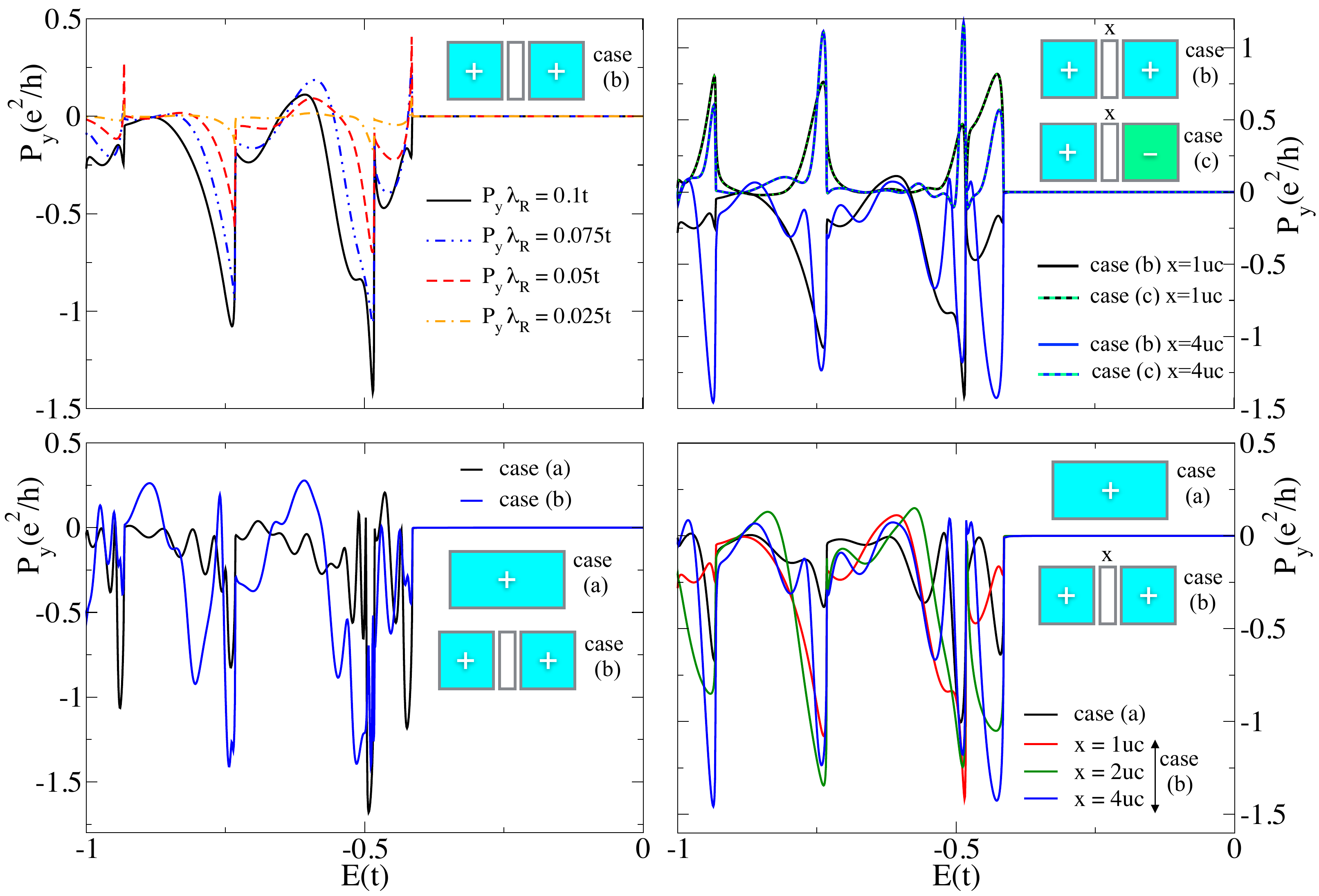}
	\caption{\small (Color online) Polarization $P_y$ of several systems based on a 11-AGNR. 
	Top left panel: system with two Rashba regions of 4 uc length with the same sign of the Rashba coupling (+R,+R) separated by a 1 uc spacer, calculated for several values of  $\lambda _R$: 0.1 $t$, 0.075 $t$ 0.05 $t$, and 0.025 $t$. 
	Bottom left panel: 11-AGNR  with a continuous central region (20 uc length) with positive Rashba SOI (black), compared to a truncated superlattice with the same total area subject to SOI, divided in five periodically repeated regions of 4 uc separated by no-Rashba 1-uc spacers (blue).
	Right panels: dependence on the spacer x. For the top panel, the central region is composed of two Rashba regions  (+R,+R) and (+R,$-$R), intercalated by a spacer with 1 and 4 uc; the bottom depicts the results for superlattices composed of six Rashba SOI regions, separated by no-Rashba 1, 2, and 4-uc spacers, and compared to a continuous region with the same Rashba area (black curve).  }
	\label{fig:4}
	\end{figure*}
\end{center}

Before we analyze how different geometric factors influence the amount of spin polarization of the current, we check first that the variation of the Rashba SOI strength does not alter our conclusions. We choose the same geometry of the previous figure, 11-AGNR, with two Rashba regions of the same sign, (+R,+R). The top left panel of Fig. \ref{fig:4} shows $P_y$ for several values of the Rashba coupling constant $\lambda _R$. It can be observed that there is a reduction of the maxima with $\lambda _R$, but it would be detectable in the instances for which proximity effects in graphene enhance the SOI interaction \cite{Calleja2015,Dedkov2008}. In any case, our conclusions are of relevance for other 2D materials besides graphene with a larger SOI.

The role of scattering by multiple interfaces between Rashba and no-Rashba parts can be studied by comparing the polarization of a single Rashba area with that with multiple SOI regions with spacers, maintaining the total area {\em and length} affected by SOI constant. 
Since the spin of the carriers precesses when moving through a Rashba region, keeping the lengths equal amounts, in principle, to having the same SOI effects in both cases \cite{Datta1990,Mireles2001}. 

We show in the bottom left panel Fig. \ref{fig:4} the polarization corresponding to an 11-AGNR formed by five repetitions of the Rashba graphene region with and without spacers (20 uc in total); we name them superlattice (blue curve) and continuous structures (black curve), respectively.
We observe that, in general,  the inclusion of more SOI/no-SOI interfaces  enhances the production of spin-polarized currents for small system sizes. Although for some specific energies the maximum value can be reached for the continuous case, in overall, the superlattice yields larger absolute values for $P_y$, as it can be clearly seen in the bottom left panel of Fig. \ref{fig:4}. 
This increasing of $P_y$ recalls the enhancement of spin polarization occurring in carbon nanotubes with equidistant impurity defects in the presence of Rashba SOI  \cite{Santos2017}. Similarly, it indicates that scattering between Rashba and no-Rashba regions is a key ingredient in the increase of the spin-polarized conductance. 


The spin-polarized current also depends on the size x of the spacers.  Results for a single spacer with different lengths (x=1 and 4 uc) are shown in  the top right panel of Fig. \ref{fig:4}
 for both (+R,+R) and (+R,$-$R) configurations. Clearly, the best responses are obtained in the case with Rashba regions of the same sign (continuous lines), L-R symmetric but with multiple SOI areas. The x=4 spacer gives the largest $P_y$, but more cases should be studied. 
This is done in the bottom right panel of Fig. \ref{fig:4}, where we display the results for a system with six repetitions of the same $+R$ region composed of 4 uc and different spacer lengths (x=1, 2, and 4 uc). The results are compared to the continuous case (black curve). The maximum value is attained for the cases with larger spacers, but it also varies very rapidly with the energy; in fact for some particular energies it can be even lower than for the no-spacer case.

\begin{center}
	\begin{figure}[h]
		 \includegraphics[width=1.0\columnwidth]{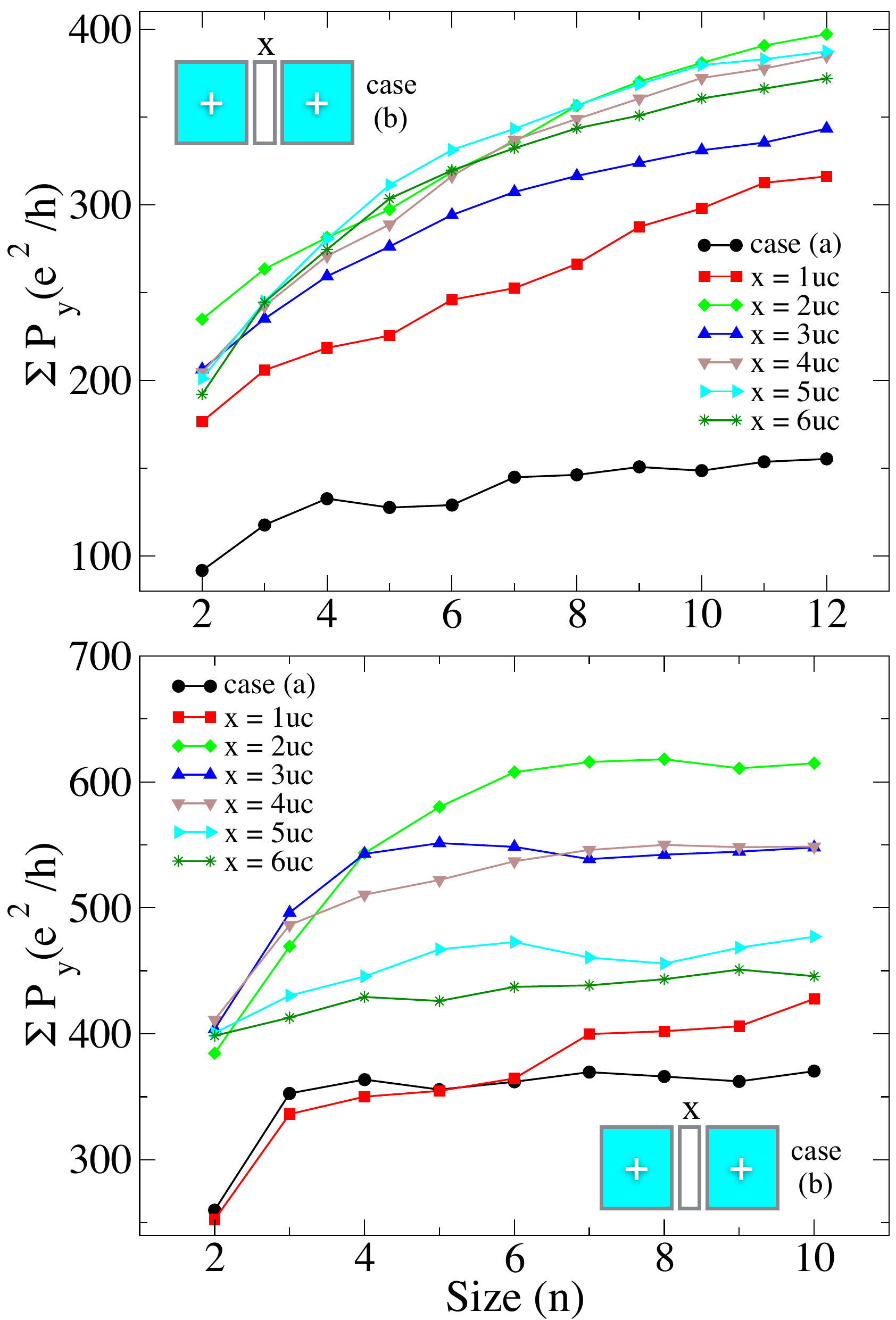}	       
		\caption{\small (Color online)  Integrated polarization values from $-1$ eV to 0 eV,  as functions of the size of the spacer area and the number of supercell repetitions. Upper and bottom panels are for 11-AGNRs and 11-ZGNRs, respectively}
		\label{fig:7}
	\end{figure}
\end{center}

As the number of spacer layers increases, many features appear in the spin polarization. This is related to the larger number of no-Rashba/Rashba interfaces and therefore the increase of the scattering. To better quantify the amount of spin polarization over an energy interval we propose another magnitude given as an integrated polarization, i.e.,

\begin{equation}
\sum P_y=\int |P_y(E)| dE\,\,, 
\end{equation}
in which the energy range for the integrated polarization was chosen to be  from zero to $-1t$.
As shown in Fig. \ref{fig:7}(a) and (b), the integrated polarization increases dramatically if  a small spacer is included (x=1 and 2 for the 11-AGNR and 11-ZGNR, respectively). It is important to remember that a single unit-cell spacer  (x=1) corresponds to no-Rashba region of different lengths for AGNR and ZGNR structures.  However, for both configurations the polarization saturates already for x $\approx 2$ in the sense that the maximum integrated polarization value is already attained for that spacer size. 
This saturation may depend on the Rashba coupling strength, as the Rashba precession length, and deserves to be explored in future investigations. 

The integrated polarization allows us to explore more easily the role of the number of repetitions $n$ in a system. With the exception of small oscillations, the general trend is an increase with the number of repetitions of the Rashba regions, tending to a constant value with larger $n$. 
Such trend is more evident in the zigzag case, but it can be also hinted for the armchair systems. Increasing $n$ should yield results closer to the limit of the superlattice, that  should obviously have a finite polarization, and hence the saturation with $n$. 

\section{Symmetry analysis}
\label{sec:symm}

To understand all the features of the conductances and polarization of 
 graphene nanoribbon superlattices we analyze the symmetries of the systems. In order to be able to explore these, we assume that all the Rashba regions in one device are of the same size, and the spacers are also identical. 
We would like to note that the spin-dependent conductance relations that we derive here  are equally valid for other planar quasi-one-dimensional materials with  multiple Rashba regions, so the interest of this analysis goes beyond graphene-based systems. 
 We show that it is necessary to take into account different symmetries in real and spin spaces in order to keep the Rashba Hamiltonian invariant in the L-R antisymmetric cases. The combination of spatial and spin symmetries provides relations between the spin-resolved conductances that allows for a full understanding of our results. 

We start from the simplest situation. For L-R symmetric systems, as those labeled {\it a, b, d} in Fig. 2, it is enough to consider the same symmetry operations in real and spin space, as shown for a nanoribbon with a single Rashba region in Ref. \cite{Chico2015}. If we take $x$ as the direction of the current and $y$ as the transversal direction, 
for a straight, two-terminal ribbon, these symmetries are $C_{2z}$, $M_x$, and $M_y$. $C_{2z}$ is a $\pi$ rotation over the $z$ axis, perpendicular to the plane of the nanoribbon. 
 Let us recall that a rotation has the same effect in real and spin variables, but a mirror reflection acts differently in real ($r$) and spin space ($s$). We use superindices $r, s$ when necessary to distinguish between these.  Thus, if $M_x$ changes the $x$ coordinate in real space from $x$ to $-x$, leaving the other two invariant, in spin space it leaves $\sigma_x$ invariant, and changes $\sigma_y\rightarrow -\sigma_y$ and $\sigma_z\rightarrow -\sigma_z$, due to the fact that spins transform as pseudovectors.  In fact, note that 
a mirror reflection in spin space is equivalent to a $\pi$ rotation: $M_x^{(s)}=C_{2x}^{(s)}$, so we have that $M_x = M_x^{(r)} \otimes M_x^{(s)}= M_x^{(r)} \otimes C_{2x}^{(s)}$. 

Besides these, we can also have other combinations of different real- and spin-space symmetries in L-R symmetric systems, but they give the same conductance relations as the former. 
Table \ref{tablesym1} shows the conductances and the corresponding spin-polarization relations generated by each real and spin symmetries ($r \otimes s$) of the L-R symmetric cases  ({\it a, b, d}, in Fig. 2). The symmetries are grouped in pairs that give exactly the same conductance relations. Notice that of each pair, one is trivial (at the bottom), in the sense that the same operation is performed in real and spin space, and the other combines two different symmetries in real and spin spaces. Therefore, for L-R symmetric systems, we could have derived the relations between spin-resolved conductances as in Ref. \cite{Chico2015}.

This is not the case for L-R antisymmetric devices.  In Table \ref{tablesym2} we display the symmetry relations of the truncated superlattice {\it c}, with an even number of Rashba regions with alternating signs. They are different from those present in the L-R symmetric systems. 

We provide below an example for the derivation of the symmetries listed in Tables \ref{tablesym1} and  \ref{tablesym2}. 
Suppose that our system has a certain spatial symmetry, neglecting the signs of the Rashba terms. 
If we apply this spatial symmetry operation to the Rashba Hamiltonian, we immediately see that a spin symmetry operation has to be considered to restore the  invariance of the Hamiltonian. 
Depending on whether we have an L-R symmetric or antisymmetric case,  
 ({\it a, b, c,} or {\it d}), this can be attained using different spin symmetry operations. 
Let us first consider L-R symmetric systems, such as cases {\it a, b}, and {\it d}. If $M^{(r)}_{x}$ holds, then $M^{(r)}_{x} \otimes C^{(s)}_{2x}$ leaves invariant the Rashba Hamiltonian. Differently, for the L-R antisymmetric systems, such as  {\it c}, (+R,$-$R), a different spin rotation symmetry $C_{2y}^{(s)}$ is required to guarantee its invariance, so the full symmetry operation is $M^{(r)}_{x} \otimes C^{(s)}_{2y}$.

 \begin{table}[t]
\renewcommand{\arraystretch}{2}
\setlength{\tabcolsep}{2pt} 
\caption{Symmetry, conductance, and spin-polarization relations for superlattices with Rashba SOI in L-R symmetric systems, as \textit{cases a, b, d}.}
\begin{center}
\begin{tabular}{|c|c|c|c|}
      \hline
   {Symmetries}   &  {Conductance} & \multicolumn{2}{c|}{Spin polarization} \\  \cline{3-4}
 ($r \otimes s$)& $(x,y,z)$  & $G_{\uparrow\uparrow}-G_{\downarrow\downarrow}$  & $G_{\uparrow\downarrow}-G_{\downarrow\uparrow}$ \\
      \hline
       \hline
        {$C^{(r)}_{2x} \otimes C^{(s)}_{2y}$} & $(x,z)$  $G^{LR}_{\sigma\sigma'} = G^{LR}_{\bar\sigma\bar\sigma'}$ & $=0$ & $=0$ \\ \cline{2-4}
    $M^{(r)}_{y} \otimes C^{(s)}_{2y}$   &  $(y)$  $G^{LR}_{\sigma\sigma'} = G^{LR}_{\sigma\sigma'}$ & $\neq0$ & $\neq0$ \\  
     \hline
      {$I^{(r)} \otimes C^{(s)}_{2z}$}   & $(x,y)$ $G^{LR}_{\sigma\sigma'} = G^{LR}_{\sigma'\sigma}$ & $\neq 0$ & $=0$ \\ \cline{2-4}
  {$C^{(r)}_{2z} \otimes C^{(s)}_{2z}$}     & $(z)$ $G^{LR}_{\sigma\sigma'} = G^{LR}_{\bar\sigma'\bar\sigma}$ &  $=0$ & $\neq0$ \\ 
     \hline
 $C^{(r)}_{2y} \otimes C^{(s)}_{2x}$ & $(y,z)$ $G^{LR}_{\sigma\sigma'} = G^{LR}_{\sigma'\sigma}$ & $\neq 0$ & $0$ \\ \cline{2-4}
  $M^{(r)}_{x} \otimes C^{(s)}_{2x}$       &  $(x)$ $G^{LR}_{\sigma\sigma'} = G^{LR}_{\bar\sigma'\bar\sigma}$ & $=0$ & $\neq0$  \\   
      \hline

\end{tabular}
\end{center}
\label{tablesym1}
\end{table}

This is best seen by taking 
the Rashba Hamiltonian in its continuum form \cite{WinklerBook}, i.e., $H_R\sim \lambda(k_x\sigma _y - k_y \sigma_x )$, where $k_i$ is the momentum component and $\sigma_i$ the spin Pauli matrices associated to the  $i=x,$ $y,$ $z$ direction. On the one hand, the mirror reflection on $x$ in real space operating on $H_R$ yields $M_{x}^{(r)} H_R \sim -\lambda(k_x \sigma_y + k_y \sigma_x)=H_R^\prime$ since $(k_x, k_y, k_z) \rightarrow (-k_x, k_y, k_z)$. On the other hand, the spin rotation operation $C_{2x}^{(s)}$ transforms spins as $(\sigma_x, \sigma_y, \sigma_z) \rightarrow (\sigma_x, - \sigma_y, - \sigma_z)$, allowing us to recover the original Rashba Hamiltonian, i. e., $C_{2x}^{(s)} H_R^\prime = H_R$. Therefore, in the L-R invariant case {\it a} the symmetry $M^{(r)}_{x} \otimes C^{(s)}_{2x}$ is present. 

The same analysis can be done for the other cases depicted in Fig. 2. Without loss of generality, we focus on the cases with two Rashba regions, L-R symmetric ($b$) and L-R antisymmetric ($c$). 
For these,  
the Hamiltonian can be divided into parts, one for each Rashba region, namely $H_R=H_R^{(1)} + H_R^{(2)}$.

Continuing with the example of the spatial mirror symmetry $M_x^{(r)}$ 
for a L-R invariant system, when the two parts have the same Rashba sign, (+R,+R) (case {\it b}), $M_x^{(r)}H_R=M_x^{(r)}H_R^{(1)} + M_x^{(r)}H_R^{(2)}$. We see that $M_x^{(r)}$ interchanges the two parts since  the Rashba couplings are equal: 
 $M_x^{(r)}H_R^{(1)}=\lambda\left ( -k_x^{(2)} \sigma_y - k_y^{(2)} \sigma_x \right ) = H^{(2\prime)}_R$ and $M_x^{(r)}H_R^{(2)}=\lambda\left ( -k_x^{(1)} \sigma_y - k_y^{(1)} \sigma_x \right )= H^{(1\prime)}_R$. The spin symmetry operation $C_{2x}^{(s)}$ is needed to leave the Hamiltonian invariant: $C_{2x}^{(s)} H^{(1\prime)}_R= H^{(1)}_R$ and $C_{2x}^{(s)} H^{(2\prime)}_R= H^{(2)}_R$, as $H^{(2)}_R = H^{(1)}_R$. Thus we have that  $M_x^{(r)} \otimes C^{(s)}_{2x} $ guarantees the invariance of  the Hamiltonian.

Now we show that for L-R antisymmetric systems as case {\it c},  a different 
 spin symmetry is needed. Assume that the Rashba sign of $H_R^{(2)}$ is negative, then $H_R=H_R^{(1)} - H_R^{(2)}$. Applying  $M_x^{(r)}$  to the Hamiltonian, 
 $M_x^{(r)}H_R^{(1)}=\lambda\left ( -k_x^{(2)} \sigma_x - k_y^{(2)} \sigma_x \right ) = H^{(2\prime)}_R$
 and 
 $-M_x^{(r)}H_R^{(2)}=-\lambda\left ( +k_x^{(1)} \sigma_x + k_y^{(1)} \sigma_x \right )= -H^{(1\prime)}_R$, 
 leading to $M_x^{(r)}H_R=H_R^{(2\prime)}-H_R^{(1\prime)}=H_R^{\prime}$.
Then, it is necessary a rotation $C_{2y}^{(s)}$  on $H^{(2\prime)}_R - H^{(1\prime)}_R$ to recover invariance, i.e.,  ($M^{(r)}_{x} \otimes C^{(s)}_{2y}) H_R = H_R$.  With the same procedure, 
all the symmetries of  the Hamiltonian  can be found. They are collected in the first column of Tables \ref{tablesym1} and \ref{tablesym2}. It is important to mention that we have used the momentum-space scenario ($k^{(1)}$ and $k^{(2)}$ wave vectors), to bring light more directly  to the symmetries involved in the systems.

\begin{table}[h]
\renewcommand{\arraystretch}{2}
\setlength{\tabcolsep}{2pt} 
\caption{Symmetry, conductance, and spin-polarization relations for superlattices with Rashba SOI in L-R antisymmetric systems, as \textit{case c}.}
\begin{center}
\begin{tabular}{|c|c|c|c|}
      \hline
   {Symmetries}   &  {Conductance} & \multicolumn{2}{c|}{Spin polarization} \\  \cline{3-4}
 ($r \otimes s$)& $(x,y,z)$  & $G_{\uparrow\uparrow}-G_{\downarrow\downarrow}$  & $G_{\uparrow\downarrow}-G_{\downarrow\uparrow}$ \\
      \hline
       \hline
        {$C^{(r)}_{2x} \otimes C^{(s)}_{2y}$} & $(x,z)$  $G^{LR}_{\sigma\sigma'} = G^{LR}_{\bar\sigma\bar\sigma'}$ & $=0$ & $=0$ \\ \cline{2-4}
    $M^{(r)}_{y} \otimes C^{(s)}_{2y}$   &  $(y)$  $G^{LR}_{\sigma\sigma'} = G^{LR}_{\sigma\sigma'}$ & $\neq0$ & $\neq0$ \\  
     \hline
  $I^{(r)}  \otimes {\mathbbm{1}}^{(s)}$   & $(x,y,z)$ $G^{LR}_{\sigma\sigma'} = G^{LR}_{\bar\sigma'\bar\sigma}$ & $ = 0$ & \setlength{\fboxrule}{3 pt}
\fcolorbox{red}{white}{\large{$\neq0$}} \\ 
  $C^{(r)}_{2z} \otimes {\mathbbm{1}}^{(s)} $  & $(x,y,z)$ $G^{LR}_{\sigma\sigma'} = G^{LR}_{\bar\sigma'\bar\sigma}$ & $ = 0$ & \setlength{\fboxrule}{3 pt}
\fcolorbox{red}{white}{\large{$\neq0$}}  \\ 
     \hline
 $C^{(r)}_{2y} \otimes C^{(s)}_{2y}$ & $(x,z)$ $G^{LR}_{\sigma\sigma'} = G^{LR}_{\sigma'\sigma}$ & $\neq 0$ & $ = 0$ \\ \cline{2-4}
  $M^{(r)}_{x} \otimes C^{(s)}_{2y}$       &  $(y)$ $G^{LR}_{\sigma\sigma'} = G^{LR}_{\bar\sigma'\bar\sigma}$ & $=0$ & \setlength{\fboxrule}{3 pt}
\fcolorbox{red}{white}{\large{$\neq0$}}  \\  
      \hline
\end{tabular}
\end{center}
\label{tablesym2}
\end{table}

The 
other 
columns
 of these Tables show the conductance relations derived from the respective symmetries. Considering the same example, namely, $M^{(r)}_{x} \otimes C^{(s)}_{2y}$, it is possible to infer that the conductance from left (with spin $\sigma$) to right lead (with spin $\sigma'$), $G^{LR}_{\sigma \sigma'}$, depends on the spin polarization direction. In this way, $G^{LR}_{\sigma \sigma'} = G^{RL}_{\bar\sigma \bar\sigma'}$ for the $x$ and $z$ spin projection directions and $G^{LR}_{\sigma \sigma'} = G^{RL}_{\sigma \sigma'}$ for the $y$ spin direction. Note that $M^{(r)}_{x}$ changes the direction of the current; therefore, time-reversal symmetry is needed, $G^{LR}_{\sigma \sigma'} = G^{RL}_{\bar\sigma' \bar\sigma}$, to derive from the previous expressions the final equalities,  namely, $G^{LR}_{\sigma \sigma'} = G^{LR}_{\sigma' \sigma}$ for the $x$ and $z$ spin directions and $G^{LR}_{\sigma \sigma'} = G^{LR}_{\bar\sigma' \bar\sigma}$ for the $y$ direction.

\begin{center}
	\begin{figure*}[t]
	\centering
		\includegraphics[width=1.75\columnwidth]{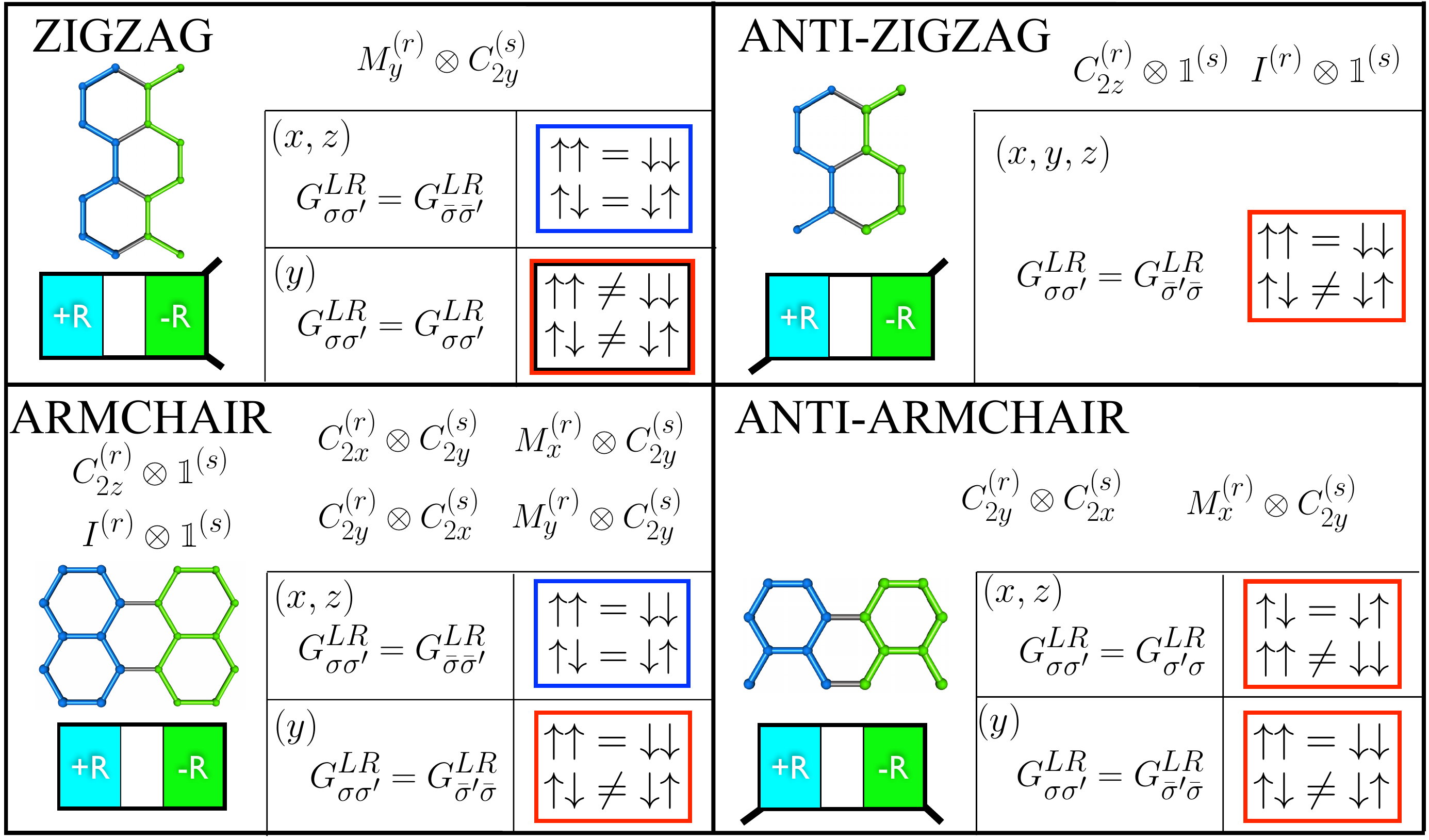}
		\caption{\small (Color online) Summary of the symmetries and spin-dependent conductance relations of L-R antisymmetric systems composed of zigzag, anti-zigzag, armchair, and anti-armchair truncated superlattices.}
		\label{fig:8}
	\end{figure*}
\end{center}

\begin{center}
	\begin{figure*}[t]
	\centering
		\includegraphics[width=1.75\columnwidth]{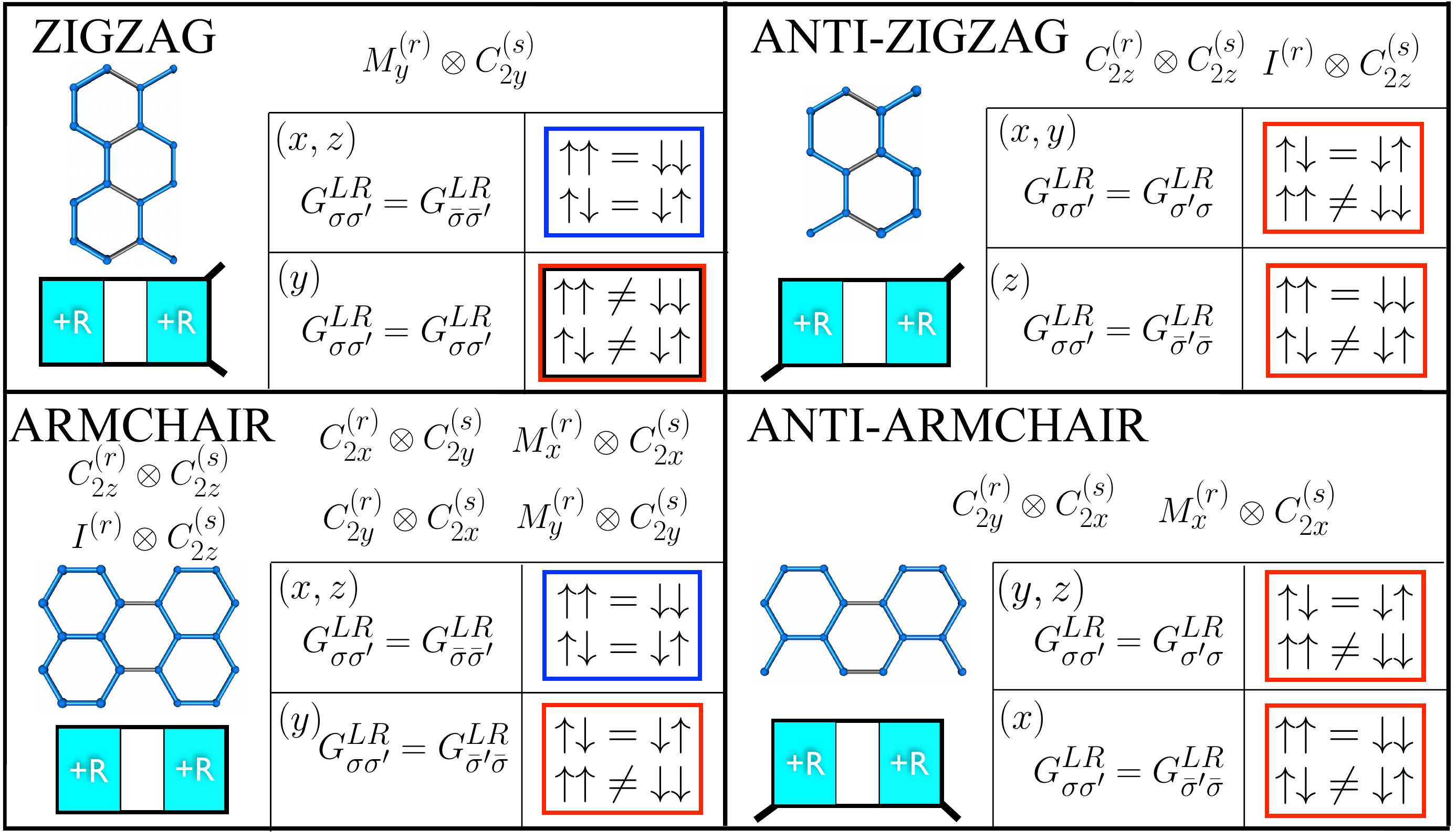}
		\caption{\small (Color online) Summary of the symmetries and spin-dependent conductance relations of L-R symmetric systems composed of zigzag, anti-zigzag, armchair, and anti-armchair truncated superlattices.}
		\label{fig:9}
	\end{figure*}
\end{center}

A summary of the predicted spin-resolved conductance relations for zigzag, anti-zigzag, armchair and anti-armchair based L-R antisymmetric systems are depicted in Fig. \ref{fig:8}, separately for the three spin directions ($x,y,z$). The relevant spatial and spin symmetry operations are indicated for each lattice. 
 Red and blue squares mark the cases with and without spin-polarized currents, respectively, and their origin, i.e., from spin-flip or spin-conserved conductance differences.  We have checked all the cases with numerical calculations. If we pay attention to the transversal spin direction ($y$), in all the studied systems the polarizations $P_y$ stem from spin-flip conductances, with the exception of the zigzag case that shows also a spin-conserved component, although this is small compared to spin-flip contribution. The polarizations in the other spin directions ($x$ and $z$) are also small compared to $y$-direction due to the geometry of the Rashba spin-orbit interaction. For completeness, we present also the summary of the symmetries and spin dependent conductance relations  of L-R symmetric systems in Fig. \ref{fig:9}, which are equal to those presented in Ref. \cite{Chico2015}.

By varying the number of corrugations of the system (or modifying the voltage gates, depending on the particular setup) the quantitative response can be tuned, but also the system can change from L-R antisymmetric to L-R symmetric. 
 This can change the symmetry properties, modifying the spin polarized current and its origin, the sign of the spin current and other characteristics, which can be exploited in the design of spintronic devices that harness this feature. For instance, a system with two gates can control the sign of the spin currents from negative (if the voltages are equal in the two gates) to positive (different voltages in the two gates); the type of polarization, spin-flip or spin-conserved can be selected, etc. Equivalently, a corrugated GNR can be mechanically modified to change the spin-polarized current, as a mechanical spin-flip switch, constituting a spin-straintronic device.

\section{Final Remarks}
\label{sec:final}

We have found that spin-polarized conductance can be enhanced in corrugated graphene nanoribbon systems, described successfully as multiple Rashba regions. The effect is larger for systems with the same sign of the Rashba SOI, 
that we have dubbed as L-R symmetric systems. The inclusion of SOI/no-SOI interfaces via small spacers (no SOI regions) enhances the effect, but the size dependence saturates soon. This indicates that multiple scattering between SOI/no-SOI regions is the reason behind the increase of spin polarization. 

Graphene systems with an even number of multiple Rashba regions with alternating signs 
call for separate spatial and spin symmetry operations in order to explain the 
relations between spin-resolved conductances. 
Although we have only presented here numerical examples with  spin projected transversal to the current, more favorable for the obtention of the maximum 
spin polarization, we have performed a complete symmetry analysis 
including all the possible directions. 

New spintronic devices can be designed by varying the induced Rashba SOI areas, i.e., changing the number of corrugations. This effect can be also produced and tuned by Rashba areas with external applied electric fields by multiple gates or by proximity phenomena with other materials. 
Importantly, the symmetry relations reported here are general and can be used to predict spin-polarized currents in other 2D materials.

\section*{Acknowledgments}

AL would like to thank the financial support of CAPES and CNPq Brazilian agencies, FAPERJ under grants E-26/101.522/2013 and E-26/202.953/2016, and the INCT de Nanomateriais de carbono. HS wants to acknowledge the support from grants of "Comunidad de Madrid" NMAT2D-CM (S2018/NMT-4511) and ADITIMAT-CM (P2018/NMT-4411) and the Spanish MINECO under No. FIS2016-80434-P. LC acknowledges the financial support of the Spanish MINECO and the European Union under Grants No. FIS2015-64654 P/MINECO/FEDER and PGC2018-097018-B-I00.

\bibliography{bib_superlt}

\end{document}